\documentclass[prb,twocolumn,amsmath,amssymb,showpacs]{revtex4}

\usepackage{graphicx}% Include figure files
\usepackage{dcolumn}% Align table columns on decimal point
\usepackage{bm}% bold math

\def\be{\begin{equation}}
\def\ee{\end{equation}}
\def\bd{\begin{displaymath}}
\def\ed{\end{displaymath}}
\def\-{\phantom{-}}

\begin{document}

\title{Generalization of polarized spin excitations for asymmetric dimeric systems}

\author{G. Houchins}
\affiliation{Department of Physics and Astronomy, James Madison University, Harrisonburg, VA 22807}
\author{J.T. Haraldsen$^*$}
\affiliation{Department of Physics and Astronomy, James Madison University, Harrisonburg, VA 22807}
\date{\today}

\begin{abstract}

Through the use of Heisenberg spin-spin interactions, we provide analytical representations for inelastic neutron scattering eigenstates and excitation cross-sections of the general $S_1$-$S_2$ spin dimeric systems. Using an exact diagonalization approach to the spin Hamiltonian, we analyze various spin coefficients to provide general representations for the neutron scattering cross-sections of two interacting spins. We also detail a generalized method for the determination of $S_z$ polarized excitations, which provide an approximation for the excitations within an applied magnetic field. These calculations provide a general understanding of the interactions between two individual or compound spin systems, which can help provide insight into condensed matter systems like molecular magnets, quantum dots, and spintronic systems, as well as particle physics investigations into quark matter and meson interactions.

~

\noindent $^*$Corresponding Author: Dr. Jason T. Haraldsen (haraldjt@jmu.edu)


\end{abstract}
\maketitle

The study of quantum nanomagnets has been expanding rapidly due to the possible technological applications for systems like molecular magnets and quantum dots due to the presence of quantum tunneling phenomena and anisotropic effects. \cite{Dag96,Bar99,DiV99,Nie00,Fur00,Bou01,Cor02,Men99,Men00,Cif01} The complete understanding of quantum excitations and the ability to detect and observe them are two critical components for the development of applications in spintronics and spin switches for quantum computing\cite{Nie00,tyag:09}.

Molecular magnets are clusters of magnetic ions that are typically isolated from long-range magnetic interactions by non-magnetic ligands \cite{whit:10,Kah93,gatt:06,tasi:04,barc:05,chab:02,dunb:07,shat:09}, and they typically have many magnetic ions like Mn$_{12}$ and V$_{15}$\cite{barc:05,chab:02}. 
%To date, the largest molecular magnet is currently Mn$_{84}$\cite{tasi:04}. 
Recently, it has been shown that many excitations within large magnetic clusters are governed by individual sub-geometry (smaller two- and three-body components) excitations\cite{hara:11PRL}. Therefore, examining the smallest components of magnetic interactions  is critical for moving forward in gaining information for the larger and more complex systems.

From an experimental point of view, there are many techniques that can be employed to characterize and measure the properties of antiferromagnetic spin systems. These include magnetic susceptibility, inelastic neutron scattering (INS), optical/Raman spectroscopy, and electron spin resonance. \cite{Squ78,Love:87,Kah93} While many of these techniques are important for the study of the bulk properties for magnetic systems, INS provides the unique ability to investigate individual excitations and examine local interactions and structural data.

Typically, discussions of magnetic clusters are limited to specific material systems\cite{amed:02,efre:02,klem:02,wald:03,tenn:97,hara:09:JPCM,hara:09:PRB}, which doesn't always provide a complete picture of the interactions being studied. Spin 1/2 clusters have been studied in great detail by a number of theoretical and experimental groups \cite{hara:05,Wha03,luba:02,cage:03a,cage:03b,kort:04}. With regards to the spin dimer, Whangbo {\it et al.} presents a detailed analysis of general excitations \cite{Wha03}; however, this work doesn't examine the changes in the inelastic neutron scattering intensities. Recently, Furrer and Waldman published a large review of symmetric $S$ magnetic clusters \cite{furr:13}. However, because mixed valency is common within molecular magnets, it is of great importance to understand not only how symmetric systems work, but also what effects are produced by spin asymmetry and what roles are played in molecular magnets. Therefore, an examination of the spin transitions for excitations of asymmetric magnetic dimer systems is needed. 

In this article, we discuss the spin excitations for the general $S_1$-$S_2$ dimer system. We evaluate these systems within the context of an isotropic Heisenberg Hamiltonian and determine analytical representations for the eigenstates. We also provide a generalized representation for the unpolarized and polarized INS structure factor for excitations from the ground state and first excited state of any spin dimer configuration. These results allow for a generalized method to determine $z$-polarized excitations, which can be observed through zero-field, or crystal field, splitting through anisotropy and/or magnetic-field splitting. Since the application of a magnetic field will split degenerate states, the unpolarized average will produce subsequent excitations, which for mixed-valence systems, will produce multiple excitations with varying intensities. Analysis of these intensity ratios allow for easy characterization of dimeric systems as well as larger molecular magnet systems. We provide detailed examples of these methodologies in the supplemental material.

%\section{Spin excitations for magnetic dimers}

To determine specific individual excitation information, including energy and momentum dependence, one needs to look towards INS. In materials with long-range magnetic ordering, the energy transfer $\omega$ is dependent on momentum transfer $\bf q$, which produces a dispersion relationship. However, for magnetic clusters, there is no dispersion, since excitations are provided in discrete quantum steps. Therefore, to gain specific information about the exchange parameters and magnetic structure, one has to look at the scattering intensity or cross-section $S({\bf q}, \omega)$ of the excitations. 

Typically, magnetic excitations produced by INS follow the selection rule of transition in S$_{tot}$ of $\pm$1 or 0. However, in cluster and molecular magnet systems, the INS excitations are specific to the spin state being excited. This modification means that neutrons can only excite energy levels of $\Delta S_{i}$ = $\pm$1 or 0, where $i$ is the specific basis state of the sub-geometries (typically dimers and trimers). Since each spin state has a specific basis associated with its excitations, the different state excitations are bound by that basis. Within a larger system, individual dimer excitations will always have the same functional form of $(1-\cos(qa))$ multiplied by a structure factor coefficient that is dependent on the specific dimer system that is being excited. Each spin state excited within the dimer determines a unique constant.

\begin{figure}
\centering
\includegraphics[width=1.0 \linewidth,clip]{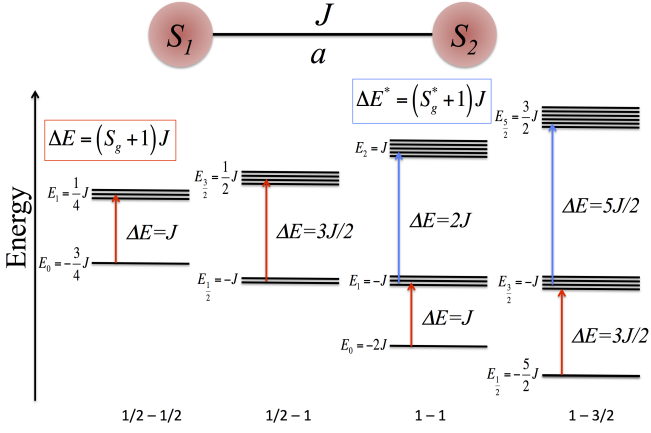} 
\caption{(Color online) An illustration of the excitations for various spin dimers detailing the ground state and first excited state excitations. Here, we define a spin dimer as the system of any two objects with definite spin separated by distance $a$ and interaction $J$. Multiple energy level line denote the number of degenerate states for the level. These will therefore split in magnetic field by $g \mu_B B S_z$.}
\label{excitations}
\end{figure}

To calculate the inelastic neutron scattering excitations for a given magnetic system, a standard Heisenberg Hamiltonian was considered to examine spin-spin exchange correlations 
%\be {\cal H} = \sum _{<\alpha \beta>} { J_{\alpha \beta}}\, \bm{S}_{\alpha}\cdot\bm{ S}_{\beta} -  \mu_B B_z \sum_{\alpha} \bm{g}_{\alpha}\, S_{z,\alpha}.
%\label{dimerH} \ee
%\noindent Here, $J_{\alpha \beta}$ is the superexchange parameter between spins $\bf{S}_{\alpha}$ and $\bf{S}_{\beta}$. The second term shows the application of a $z$-axis external magnetic field $B_z$, where $\bf{g}_{\alpha}$ is the gyromagnetic tensor for each specific magnetic ion, $\mu_B$ is the Bohr magneton, and $S_z$ is the $z$-component of the spin operator. From the Hamiltonian, the energy eigenstates and eigenvalues may be found by diagonalizing the magnetic Hamiltonian on a convenient basis. The usual set of $\hat z$-polarized magnetic basis states would then be employed.
for a generalized spin dimer of $S_1$-$S_2$. As shown in Figure \ref{excitations}, the spin dimer consists of two objects with distinct spins that interact through the isotropic Heisenberg interaction, where the Hamiltonian can be written as
\be {\cal H} = { J}\, \bm{ S}_{1}\cdot\bm{S}_{2} - \mu_B B_z \big( \bm{g}_1  S_{z,1} + \bm{g}_2 S_{z,2}),
\label{dimerH} \ee
\noindent where $J$ is the superexchange parameter between spins $\bf{S}_{1}$ and $\bf{S}_{2}$. The second term shows the application of a $z$-axis external magnetic field $B_z$, where $\bf{g}$ is the gyromagnetic tensor for each specific magnetic ion, $\mu_B$ is the Bohr magneton, and $S_z$ is the $z$-component of the spin operator. From the Hamiltonian, the energy eigenstates and eigenvalues may be found by diagonalizing the magnetic Hamiltonian on a convenient basis. The usual set of $\hat z$-polarized magnetic basis states would then be employed.From the Clebsch-Gordon series, the spin decomposition of the general spin dimer is given as
\be 
\textbf{S} \otimes \textbf{S} = \sum_0^S \textbf{S} .
\ee
\noindent Therefore, any magnetic state will have $2S_{tot}+1$ degenerate states and overall $(2S_1+1)(2S_2+1)$ total states. The zero-field energy eigenstates can be shown as a function of the $S_{tot}$ of that state and $S_1$ and $S_2$ as given by
\begin{multline}
E_{S_{tot},S_1,S_2} = \frac{1}{2}J \Big(S_{tot}\big(S_{tot}+1\big)-\\
S_{1}\big(S_{1}+1\big)-S_{2}\big(S_{2}+1\big)\Big)-E_{Zee},
\end{multline}
where $E_{Zee}$ = $g_{S_z}\mu_B B S_z$, and $g_{S_z}$ is a linear combination of the local $\bm{g}$ tensors.\cite{Kah93} As illustrated in Figure \ref{excitations}, any excitation from a $S_{tot}$ state in the dimer is given by
\be 
\Delta E = \left(S_{tot}+1\right)J - \Delta E_{Zee}.
\label{groundE}
\ee

\noindent Once the energy eigenstates of the system have been determined, the inelastic neutron scattering cross section can be calculated by evaluating the inelastic structure factor $S(\bm{q},\omega)$. 

%\section{Polarized structure factor coefficients}

In ``spin-only" magnetic neutron scattering, the differential cross-section for the inelastic scattering of an incident neutron from a magnetic system in an initial state $\Psi_{i}$, with momentum transfer $\hbar \bm{q}$ and energy transfer $\hbar\omega$, is given by
\be 
\dfrac{d^2\sigma}{d\Omega d\omega}=R(\bm{q})\sum\limits_{ba}\left(\delta_{ba} - \frac{q_b q_a}{q^2}\right)  S_{ba}^{(fi)}(\bm{q}\,,\omega ) 
\ee
where 
\be 
R(\bm{q})=(\gamma r_0)^2 \dfrac{k'}{k}e^{-2W(\bm{q})}.
\ee
In the equation above, $\gamma$ = 1.91, $r_0$ is the classical electron radius, and $e^{-2W(\bm{q})}$is the Debye-Waller factor\cite{furr:13,Love:87}. For transitions between discrete energy levels, the standard time integral gives a trivial delta function in the energy transfer. Therefore, for clusters excitations, the energy component can be pulled out and the structure factor becomes dependent only on ${\bf q}$. Therefore, we define polarized neutron scattering structure factor
\begin{multline}
S_{ba}^{(fi)}(\bm{q}) =
\sum\limits_{\lambda_i, \lambda_f}p_\lambda \langle \Psi_i (\lambda_i) | V_b^{\dagger} | \Psi_f (\lambda_f)\rangle\ \\ 
\langle \Psi_f (\lambda_f)| V_a | \Psi_i(\lambda_i) \rangle 
\end{multline}
\noindent where the vector $V_a(\bm{q}\,) $ is a sum of spin operators over all magnetic
ions in a unit cell, 
\be 
V_a = \sum_{_i} F_{i}(\bm{q}){S}_a(\bm{x}_i)\;
e^{i\bm{q} \cdot \bm{x}_i } \ ,
\label{Va_defn} 
\ee
\noindent Here, $F_{i}(\bm{q})$ is the normalized spin density for each magnetic moment and $p_\lambda$ is the thermal population factor which is described by
\be 
p_\lambda = \dfrac{1}{Z}e^{-\frac{E_\lambda}{k_B T}}
\ee
This provides a temperature dependence on the magnetic excitations. For simplicity, we can consider excitations at $T$ = 0.

These polarized structure factors describe the excitations for specific $z$-polarized transitions. However, in the case of a magnetic system with no energy splitting, it is necessary to take an unpolarized average of the polarized bases, which will give the unpolarized structure factor, $\langle S(\bm{q}) \rangle$. This is an average over the polarized transitions and not a sum due to the probably of populated excitations.

\begin{figure}
\centering
\includegraphics[width=1.0 \linewidth,clip]{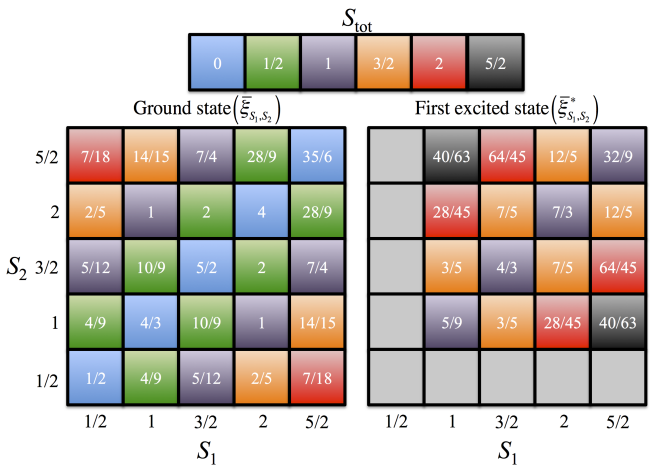} 
\caption{(Color online) The INS structure factor coefficients grid for the ground state and first excited state excitations in general $S_1$-$S_2$ dimer. The colored diagonals denote the various spin ground states from $S_{tot}$ = 0 - 5/2.}
\label{grid}
\end{figure}

The aforementioned result applies to neutron scattering from single crystals. However, many molecular magnetic are actually in powder form. Therefore, to interpret neutron experiments on powder samples, we require an orientation average of the unpolarized single-crystal  neutron scattering structure factor. We define this powder average by 
\be 
{\bar S}(q) = \int \frac{d\Omega_{\hat q}}{4\pi}\, S(\bm{q}\,) \ . 
\label{Strpowavg} 
\ee
For any dimer excitation, the structure factor always takes on the functional form of
\be 
S(\bm{q}) = \bar \xi_{S_1,S_2}  \Big(\frac{F_1^2({\bf q})+F_2^2({\bf q})}{2}- F_1({\bf q})F_2({\bf q})\cos (\bm{q}\cdot \Delta \bm{r})\Big),
\ee 
where $F_1({\bf q})$ and $F_2({\bf q})$ are magnetic form factors for each ion and $\bar \xi_{S_1,S_2} $ is a unpolarized spin coefficient that is described by averaging the polarized spin correlations $\xi_{|S_1,S_2,S_{tot}^i,S_{x}^i\rangle}^{|S_1,S_2,S_{tot}^f,S_{z}^f\rangle}$ between all possible initial and final states. If one assumes a symmetric dimer, then the equation is simplified to 
\be 
S(\bm{q}) = \bar \xi_{S_1,S_2} F^2({\bf q}) \big(1-\cos (\bm{q}\cdot \Delta \bm{r})\big).
\ee 
To find the powered averaged structure factor, the integration over all angles simply change $\cos (\bm{q}\cdot \Delta \bm{r})$ to $j_0(qa)=\sin(qa)/qa$, where $a$ is the distance between the two interacting spin systems, $q$ is the magnitude of the momentum transfer.
%\be 
%{\bar S}(q)= \bar \xi_{S_1,S_2}  \big(\frac{F_1^2(q)+F_2^2(q)}{2}- F_1(q)F_2(q) j_0(qa))\big),
%%\ee 
%where $a$ is the distance between the two interacting spin systems, $q$ is the magnitude of the momentum transfer, and $j_0(x)=\sin(x)/x$.

%To determine the coefficient of the structure factor for each possible excitation, we must calculate the probability for each transition from the initial to final spin states. This provides the polarized transitions, which are then average together to get the unpolarized structure factor. These coefficients make up a spin transition matrix (similar to that in Table \ref{ex-table}.) Here, the matrix contains the information needed for any polarized transition from the $|S_{tot}^i,S_{z}^i\rangle$ state to the $\langle S_{tot}^f,S_{z}^f|$ state for a specific dimeric system. Because of specific symmetries, only the $\xi^{1,1}$ entry is needed to calculate the unpolarized spin correlation coefficient, 
%\begin{multline}
%\bar \xi_{S_1,S_2}  = \langle S_{tot}^f,S_{z}^f|V^+|S_{tot}^i,S_{z}^i\rangle^2
%\frac{2S_{tot}^f+1}{3\big(2S_{tot}^i+1\big)} \\= \xi^{1,1}\frac{2S_{tot}^f+1}{3\big(2S_{tot}^i+1\big)}
%\end{multline}

However, by evaluating and examining the polarized matrix coefficients for various dimer combinations and using various pattern recognition techniques, an analytical solution for any excitation from the spin ground state $S_g=S_1-S_2$ can be deduced. Therefore, the unpolarized structure factor coefficient can be given by
%\be \bar \xi_{S_1,S_2} = \frac{1}{6}\frac{(S_1+S_2-S_g)(S_1+S_2+S_g+2)}{S_g+1}. \ee
%This can be further reduced to
\be \bar \xi_{S_1,S_2} = \frac{2}{3}\frac{S_2(S_1+1)}{S_1-S_2+1}, \ee
which provides a generalized formula for any spin combination of $S_1$ and $S_2$ using the above expression for $S_g$ and assuming that $S_1\geq S_2$. Furthermore, this methodology allows us to calculate the neutron scattering coefficients for the first excited state, $S_g^*=S_1-S_2+1$. However, the equation varies slightly and can be written as
%\be 
%\bar \xi_{S_1,S_2}^* = \frac{4}{6}\frac{S_g^*(S_1+S_2-S_g^*)(S_1+S_2+S_g^*+2)}{(S_g^*+1)(2S_g^*+1)}.
%\ee
\be 
\bar \xi_{S_1,S_2}^* = \frac{2}{3}\frac{(S_1-S_2+1)(2S_1+3)(2S_2-1)}{(S_1-S_2+2)(2(S_1-S_2)+3)}.
\ee
Therefore, dimer excitations from the ground state or the first excited state can be easily determined.

Figure \ref{grid} shows the unpolarized neutron scattering coefficient $\bar \xi_{S_1,S_2}$ as a function of $S_1$ and $S_2$ for excitations from the ground state and the first excited state. Here, the colors of the diagonals indicate the initial total spin state for each transition. These values allow for easy determination of dimeric excitations and provides a better understanding of how the individual spins of the dimer dictate the overall intensity. It is shown that symmetric dimers $S_1$=$S_2$ have the largest intensities and deviations from the symmetric configuration will decrease your overall structure factor coefficient. Therefore, if you are examining molecular magnets that consists of multiple dimer configurations (symmetric and asymmetric), then an analysis of the intensity difference can distinguish between them. 

Figure \ref{grid} also shows the structure factor coefficients for the excitations from the first excited state, $\xi_{S_1,S_2}^*$, where the excited may have been thermally or electrically populated. Analysis of excited state coefficients can provide increased understanding and characterization of exchange interactions. 

\begin{figure}
\centering
\includegraphics[width=1.0 \linewidth,clip]{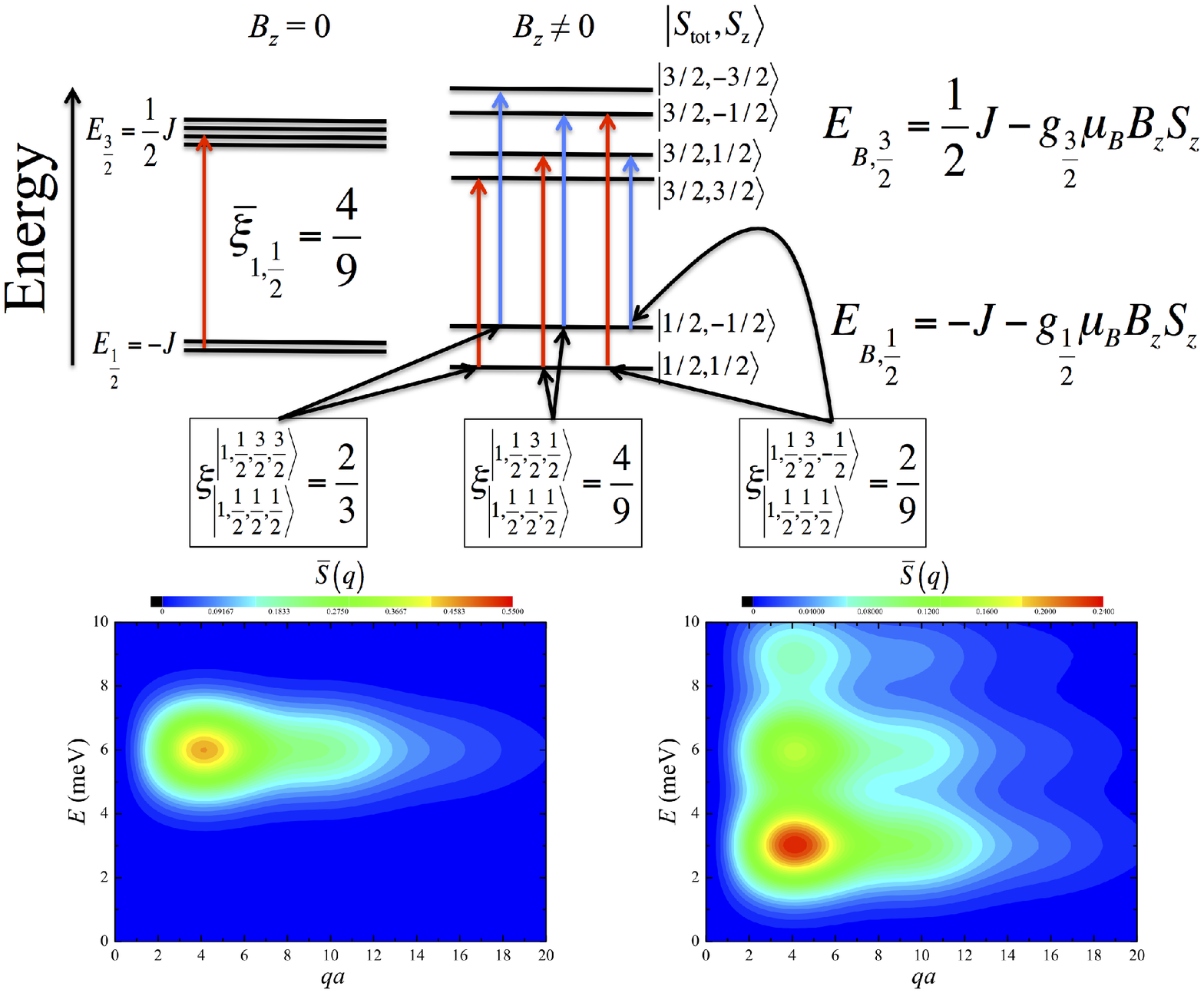} 
\caption{(Color online) The degenerate energy levels of the 1-1/2 isotropic dimer can be split through the application of an external magnetic field. Here, the spin-1/2 ground state and spin-3/2 excited states shift by the standard Zeeman energy, where $g_{\frac{3}{2}}$ and $g_{\frac{1}{2}}$ denote the mixed valence gyromagnetic tensor constants. The two lower panels show the change in the inelastic neutron scattering spectra for (left) zero field and (right) an applied field on a simulated V$^{3+}$-V$^{4+}$ dimer (spin 1 and spin 1/2), where $J$ = 4.0 meV, $\Delta g \mu_B B_z$ = 1.0 meV\cite{particle}. The magnetic form factor produces a drop off of intensity, but the main peaks remain for comparison.
}
\label{split}
\end{figure}

While having the unpolarized transition coefficients are useful for understanding the trends and overall describing isotropic dimer excitations, they provide only information for those single excitations, which are typically normalized in inelastic neutron scattering measurement.  However, these highly degenerate states are typically split either through the application of a magnetic field (shown in Fig. \ref{split}) or an internal zero-field anisotropy through crystal-field interactions. When these interactions break energy-level degeneracy, then the individual transitions become apparent. 

To determine the coefficient of the structure factor for each possible excitation, we must calculate the probability for each transition from the initial to final spin states, which provides the polarized transitions that are averaged together to get the unpolarized structure factor. These coefficients make up a spin transition matrix (similar to that in Table \ref{ex-table}.) Here, the matrix contains the information needed for any polarized transition from the $|S_{tot}^i,S_{z}^i\rangle$ state to the $\langle S_{tot}^f,S_{z}^f|$ state for a specific dimeric system. Because of specific symmetries, only the $\xi^{1,1}$ entry is needed to calculate the unpolarized spin correlation coefficient, 
\begin{multline}
\bar \xi_{S_1,S_2}  = \langle S_{tot}^f,S_{z}^f|V^+|S_{tot}^i,S_{z}^i\rangle^2
\frac{2S_{tot}^f+1}{3\big(2S_{tot}^i+1\big)} \\
= \xi^{1,1}\frac{2S_{tot}^f+1}{3\big(2S_{tot}^i+1\big)}
\end{multline}

\begin{table}[t]
\caption{Correlation space for excitations between initial and final spin states. The polarized spin correlations coefficients are shown as $\xi_{|S_1,S_2,S_{tot}^i,S_{z}^i\rangle}^{|S_1,S_2,S_{tot}^f,S_{z}^f\rangle}$ = $\xi^{m,n}$, where $m$ and $n$ denote the corresponding row and column.}
\begin{tabular}{|c|c|c|c|c|}
\hline
$S_1$,$S_2$ & $|S_{tot}^i,S_{z}^i\rangle$ & $|S_{tot}^i,S_{z}^i-1\rangle$  & $\cdots$ & $|S_{tot}^i,-S_{z}^i\rangle$ \\
\hline
$\langle S_{tot}^f,S_{z}^f|$ & $\xi^{1,1}$ & $\xi^{1,2}$  & $\cdots$ & $\xi^{1,2S_{tot}^i+1}$\\
\hline
$\langle S_{tot}^f,S_{z}^f-1|$ & $\xi^{2,1}$ & $\xi^{2,2}$ & $\cdots$ & $\xi^{2,2S_{tot}^i+1}$\\
%\hline
%$\langle S_{tot}^f,S_{z}^f-2|$ & & & $\cdots$ & X\\
%\hline
%$\langle S_{tot}^f,S_{z}^f-3|$ & X & & $\cdots$ & X\\
\hline
$\vdots$ & $\vdots$ & $\vdots$ & & $\vdots$\\
\hline
$\langle S_{tot}^f,-S_{z}^f|$ & $\xi^{2S_{tot}^f+1,1}$ & $\xi^{2S_{tot}^f+1,1}$ & $\cdots$ & $\xi^{2S_{tot}^f+1,2S_{tot}^i+1}$\\
\hline
\end{tabular}
\label{ex-table}
\end{table}

Therefore, it is useful to be able to determine the polarized matrix coefficients that describe individual transition coefficients (illustrated in Fig. \ref{split}). Through an analysis of these transitions, the intensity ratios of the split excitations can be used to distinguish and characterize the specific dimer being excited. Typically, these coefficients require a detailed analysis of the eigenstates within the dimer system. However, due to specific symmetries in the coefficient matrix, it is possible to reproduce all transition coefficients from the only the first coefficient value and the unpolarized average. 
%The correlation matrix can therefore be defined through by the symmetry
This is described in detail in the supplemental material.
%\be
%\xi_{|S_1,S_2,S_{tot}^i,S_{z}^i\rangle}^{|S_1,S_2,S_{tot}^f,S_{z}^f\rangle} = \xi_{|S_1,S_2,S_{tot}^i,-S_{z}^i\rangle}^{|S_1,S_2,S_{tot}^f,-S_{z}^f\rangle}, 
%\ee
%which provides a mirror symmetry to the asymmetric matrix. In addition to this rule, the sum of all row values is given by
%\be
%\sum_{i=1}^{2S_{tot}^i+1} \xi^{1,i} = \xi^{1,1}
%\ee
%and the sum of all column values gives
%\be
%\sum_{i=1}^{2S_{tot}^f+1} \xi^{i,1} = 3 \bar \xi_{S_1,S_2}
%\ee
%Finally, the diagonal values provide another constraint, which can be written as
%\be
%\sum_{1}^{2S_{tot}^i+1} \xi^{i,i} = (2S_{tot}^i+1)\bar \xi_{S_1,S_2}.
%\ee
Using these symmetries, one can produce a system of linear equations for the determination of all transition values of $\xi^{m,n}$ when given the one or two of the diagonal values ($\xi^{1,1}$ and $\xi^{2,2}$) and the unpolarized average, which is dependent only on $S_1$ and $S_2$. This methodology is discussed in more detail in the supplementary material, where we provide specific examples for the determination of the coefficients for two spin dimer combinations.

With the polarized coefficients known, individual excitations can be characterized through zero-field or magnetic-field splitting. Figure \ref{split} shows the degenerate energy levels of the 1-1/2 isotropic dimer being split through the application of an external magnetic field. Here, the single transition is split into six separate transitions as $S_{g}$ = 1/2 doublet and  $S_{g}^*$ = 3/2 quartet are shifted by the applied field ($E_{Zee}$). Therefore, the unpolarized average coefficient of $\bar \xi_{1,1/2}$ = 4/9, will split into three polarized coefficients of 2/3, 4/9, and 2/9. This splitting will impact the observed neutron scattering excitations, which is shown in the lower panels of Fig. \ref{split}, where an analysis of the INS intensities will allow for the characterization of the spin dimers by looking at the ratios of the individual transition coefficients.

In conclusion, we provide a detailed understanding of the energy eigenstates and inelastic neutron scattering structure factors for the ground state and first excited state excitations of the generalized $S_1$-$S_2$ spin dimer. We determine an analytic form for the unpolarized structure factors, as well as produced a general methodology for the determination of the individual polarized structure factors, which are needed for systems that have non-degenerate energy levels in the case of anisotropy or the application of a magnetic field. Overall, these calculations can help push the identification and characterization of magnetic systems like molecular magnets and quantum nanostructures and dots.

While these calculations are performed within the context of condensed matter magnetic systems, the general context and methodology of spin excitations for two interacting spins may be useful for the understanding spin excitations in quark matter; particularly meson excitations\cite{oste:92,krew:80}. The generalization of the spin interactions is independent of probing source and simply investigates the general operations on a local moment.

\section*{Acknowledgements}
We would to thank C. Constantin, T. Ahmed, and J.-X. Zhu for useful and insightful discussions. This project was supported by the Department of Physics and Astronomy at James Madison University.

\end{document}

\title{Generalization of polarized spin excitations for asymmetric dimeric systems: Supplemental material}

\author{G. Houchins}
\affiliation{Department of Physics and Astronomy, James Madison University, Harrisonburg, VA 22807}

\author{J.T. Haraldsen}
\affiliation{Department of Physics and Astronomy, James Madison University, Harrisonburg, VA 22807}

\begin{abstract}

In the supplementary material, we provide a clarification of the structure factor coefficient matrix and its respective symmetries. Furthermore, we show specific examples of how one can use the coefficient matrix to find not only the unpolarized average, but the polarized coefficients as well. Examples are done for a spin 1 - spin 1/2 dimer and a spin 5/2 - spin 1/2 dimer.

\end{abstract}
\maketitle

\section{Dimer Coefficient Table}

To determine the coefficient of the structure factor for each possible excitation, we must calculate the probability for each transition from the initial to final spin states. This provides the polarized transitions, which are then average together to get the unpolarized structure factor. These coefficients make up a spin transition matrix (similar to that in Table \ref{ex-table}.) Here, the matrix contains the information needed for any polarized transition from the $|S_{tot}^i,S_{z}^i\rangle$ state to the $\langle S_{tot}^f,S_{z}^f|$ state for a specific dimeric system. Because of specific symmetries, only the $\xi^{1,1}$ entry is needed to calculate the unpolarized spin correlation coefficient, 
\begin{equation}
\bar \xi_{S_1,S_2}  = \langle S_{tot}^f,S_{z}^f|V^+|S_{tot}^i,S_{z}^i\rangle^2
\frac{2S_{tot}^f+1}{3\big(2S_{tot}^i+1\big)} = \xi^{1,1}\frac{2S_{tot}^f+1}{3\big(2S_{tot}^i+1\big)}
\end{equation}

While having the unpolarized transition coefficients are good at describe isotropic dimers, the highly degenerate states can be split either through the application of magnetic field or internal zero-field anisotropy. When these energies break energy level degeneracy, the individual transitions become apparent. Therefore, it is useful to be able to easily determine the polarized matrix coefficients. Due to specific symmetries in the coefficient matrix due to the Clebsch-Gordon coefficients, it is possible to reproduce all transition coefficients from the only a few coefficient values. Specifically, 
\be
\xi_{|S_1,S_2,S_{tot}^i,S_{z}^i\rangle}^{|S_1,S_2,S_{tot}^f,S_{z}^f\rangle} = \xi_{|S_1,S_2,S_{tot}^i,-S_{z}^i\rangle}^{|S_1,S_2,S_{tot}^f,-S_{z}^f\rangle}, 
\ee
which provides a mirror symmetry to the asymmetric matrix. In addition to this rule, the sum of all row values is given by
\be
\sum_{i=1}^{2S_{tot}^i+1} \xi^{1,i} = \xi^{1,1}
\ee
and the sum of all column values gives
\be
\sum_{i=1}^{2S_{tot}^f+1} \xi^{i,1} = 3 \bar \xi_{S_1,S_2}
\ee
Finally, the diagonal values provide another constraint, which can be written as
\be
\sum_{1}^{2S_{tot}^i+1} \xi^{i,i} = (2S_{tot}^i+1)\bar \xi_{S_1,S_2}.
\ee
With these symmetries, one can produce a system of linear equations for all transition values of $\xi^{m,n}$ given the diagonal values.

\begin{table}
\caption{Correlation space for excitations between initial and final spin states. The polarized spin correlations coefficients are shown as $\xi_{|S_1,S_2,S_{tot}^i,S_{z}^i\rangle}^{|S_1,S_2,S_{tot}^f,S_{z}^f\rangle}$ = $\xi^{m,n}$, where $m$ and $n$ denote the corresponding row and column.}
\begin{tabular}{c|c|c|c|c}
$S_1$,$S_2$ & $|S_{tot}^i,S_{z}^i\rangle$ & $|S_{tot}^i,S_{z}^i-1\rangle$  & $\cdots$ & $|S_{tot}^i,-S_{z}^i\rangle$ \\
\hline
$\langle S_{tot}^f,S_{z}^f|$ & $\xi^{1,1}$ & $\xi^{1,2}$  & $\cdots$ & $\xi^{1,2S_{tot}^i+1}$\\
\hline
$\langle S_{tot}^f,S_{z}^f-1|$ & $\xi^{2,1}$ & $\xi^{2,2}$ & $\cdots$ & $\xi^{2,2S_{tot}^i+1}$\\
%\hline
%$\langle S_{tot}^f,S_{z}^f-2|$ & & & $\cdots$ & X\\
%\hline
%$\langle S_{tot}^f,S_{z}^f-3|$ & X & & $\cdots$ & X\\
\hline
$\vdots$ & $\vdots$ & $\vdots$ & & $\vdots$\\
\hline
$\langle S_{tot}^f,-S_{z}^f|$ & $\xi^{2S_{tot}^f+1,1}$ & $\xi^{2S_{tot}^f+1,1}$ & $\cdots$ & $\xi^{2S_{tot}^f+1,2S_{tot}^i+1}$\\
\hline
\end{tabular}
\label{ex-table}
\end{table}

\section{Dimer Coefficient Examples}

In this supplemental section two specific mixed valence dimer situations, $S_1=1$, $S_2=\frac{1}{2}$ and  $S_1=\frac{5}{2}$, $S_2=\frac{1}{2}$, are discussed in further detail to elaborate the procedure of determine the polarized coefficients. Determination of the average, non-polarized, coefficient, the use of symmetries of the correlation coefficients, and the implementation of linear equations are all explored through these two examples.

As discussed earlier, the sum of each of the rows of the correlation matrix add to the same number. The top row, however, contains only one value and therefore must be the value the rows add to. Using this information, the value of the first entry, $\xi^{1,1}$ times the number of rows, $2S_{tot}^f+1$ must equal the sum of all the polarized coefficients. To find the average, not that each column contains 3 and only 3 correlation coefficients.  Therefore there are 3 times the number of columns or $3\big(2S_{tot}^i+1\big)$ coefficients. Hence: 

\be
\bar \xi_{S_1,S_2}  = \langle S_{tot}^f,S_{z}^f|V^+|S_{tot}^i,S_{z}^i\rangle^2
\frac{2S_{tot}^f+1}{3\big(2S_{tot}^i+1\big)} \\= \xi^{1,1}\frac{2S_{tot}^f+1}{3\big(2S_{tot}^i+1\big)}
\ee

%\be
%\bar{\xi}_{S_1,S_2} = \langle S_{fin},S_{fin}^{tot}|V^+|S_{int}^{tot},S_{int}\rangle \langle S_{int},S_{int}|V^-|S_{fin},S_{fin}\rangle\frac{2S_{fin}+1}{2S_{int}+1}
%\ee

\begin{table}
\caption{Necessary starting coefficients, by spins, to determine the correlation space matrix}
\begin{tabular}{|c|c|c|c|c|}
\hline
$S_1$ & $S_2$ & $S_g$ & $\overline{\xi}_{S_1,S_2}$ & $\xi^{ii}$\\
\hline
$\frac{1}{2}$ & $\frac{1}{2}$ & 0 & $\frac{1}{2}$ & $\frac{1}{2}$\\
\hline
1 & $\frac{1}{2}$ & $\frac{1}{2}$ & $\frac{4}{9}$ & $\frac{2}{3}$\\
\hline
$\frac{3}{2}$ & $\frac{1}{2}$ & 1 & $\frac{5}{12}$ & $\frac{3}{4}$\\
\hline
2 & $\frac{1}{2}$ & $\frac{3}{2}$ & $\frac{2}{5}$ & $\frac{4}{5},\frac{12}{25}$\\
\hline
$\frac{5}{2}$ & $\frac{1}{2}$ & 2 &  $\frac{7}{18}$ & $\frac{5}{6},\frac{5}{9}$\\
\hline
$1$ & $1$ & 0 &  $\frac{4}{2}$ & $\frac{4}{3}$\\
\hline
$\frac{3}{2}$ & $1$ &  $\frac{10}{9}$ & $\frac{1}{2}$ & $\frac{5}{3}$\\
\hline
$2$ & $1$ & 1 &  $1$ & $\frac{9}{5}$\\
\hline
$\frac{5}{2}$ & $1$ & $\frac{3}{2}$ &  $\frac{14}{15}$ & $\frac{28}{15},\frac{4}{7}$\\
\hline
$\frac{3}{2}$ & $\frac{3}{2}$ & 0 &  $\frac{5}{2}$ & $\frac{5}{2}$\\
\hline
$2$ & $\frac{3}{2}$ & $\frac{1}{2}$ & 2 & $\frac{6}{2}$\\
\hline
$\frac{5}{2}$ & $\frac{3}{2}$ & 1 &  $\frac{7}{4}$ & $\frac{63}{20}$\\
\hline
$2$ & $2$ & 0 & 4 & $4$\\
\hline
$\frac{5}{2}$ & $2$ & $\frac{1}{2}$ &  $\frac{28}{9}$ & $\frac{14}{3}$\\
\hline
$\frac{5}{2}$ & $\frac{5}{2}$ & 0 &  $\frac{35}{6}$ & $\frac{35}{6}$\\
\hline
\end{tabular}
\label{coeff}
\end{table}

For the case of $S_1=1$ and $S_2=\frac{1}{2}$ only the value of $\xi^{1,1}$ and the average $\overline{\xi}$, which can be deduced from $\xi^{1,1}$ is required to find all polarized constant. From table \ref{coeff}, we find that $\xi^{1,1}=\frac{2}{3}$ and $\overline{\xi}=\frac{4}{9}$.

\begin{table}
\caption{Correlation space for $S_1=1$, $S_2=\frac{1}{2}$ with initial conditions and unknown variables.}
\begin{tabular}{c|c|c|}
 & $\langle \frac{1}{2},\frac{1}{2}|$ & $\langle \frac{1}{2},-\frac{1}{2}|$ \\
\hline
$\langle \frac{3}{2},\frac{3}{2}|$  &$\xi^{1,1}=\frac{2}{3}$ & 0\\
\hline
$\langle  \frac{3}{2},\frac{1}{2}$ & $\xi^{2,1}$ & $\xi^{3,1}$ \\
\hline
$\langle  \frac{3}{2},-\frac{1}{2}$ & $\xi^{3,1}$ & $\xi^{2,1}$\\
\hline
$ \langle \frac{3}{2},-\frac{3}{2}|$ & 0 & $\xi^{1,1}=\frac{2}{3}$ \\
\hline
\end{tabular}
\label{case1incomplete}
\end{table}

As shown in \ref{case1incomplete}, the symmetry 
\be
\xi_{|S_1,S_2,S_{tot}^i,S_{z}^i\rangle}^{|S_1,S_2,S_{tot}^f,S_{z}^f\rangle} = \xi_{|S_1,S_2,S_{tot}^i,-S_{z}^i\rangle}^{|S_1,S_2,S_{tot}^f,-S_{z}^f\rangle}, 
\ee
can be used to deduce that 
\be
\xi^{1,1}=\xi^{4,2}=\frac{2}{3},~
\xi^{2,1}=\xi^{3,2},~
\xi^{3,1}=\xi^{2,2}
\ee

Additionally, the sum of all column values given by 
\be
\sum_{i=1}^{2S_{tot}^f+1} \xi^{i,1} = 3 \bar \xi_{S_1,S_2}
\ee

allows us to set up the equation
\be
\frac{2}{3}+\xi^{2,1}+\xi^{3,1}=\frac{4}{3}
\ee
Due to symmetry, there is only one unique equation that can be determined from this conditon. The sum of all row values given by
\be
\sum_{i=1}^{2S_{tot}^i+1} \xi^{1,i} = \xi^{1,1}
\ee

allows us to set up the equation

\be
\xi^{2,1}+\xi^{3,1}=\frac{2}{3}
\ee

where we have substituted $\xi^{2,2}$ for $\xi^{3,1}$. Again because of symmetry there is only one equation that can be determined from this constraint. The final condition on the values of the diagonals provides no further information since there are 2 unique equations and only 2 unique unknowns. Solving the above system of two equations yields

$$\xi^{2,1}=\xi^{3,2}=\frac{4}{9}~\text{and}~ \xi^{2,2}=\xi^{3,1}=\frac{2}{9}$$.

\begin{table}
\caption{Completed correlations space for $S_1=1$, $S_2=\frac{1}{2}$}
\begin{tabular}{c|c|c|}
%\hline
 & $\langle \frac{1}{2},\frac{1}{2}|$ & $\langle \frac{1}{2},-\frac{1}{2}|$ \\
\hline
$\langle \frac{3}{2},\frac{3}{2}|$ &$\frac{2}{3}$ & 0\\
\hline
$\langle  \frac{3}{2},\frac{1}{2}$ & $\frac{4}{9}$ & $\frac{2}{9} $ \\
\hline
$\langle  \frac{3}{2},-\frac{1}{2}$ & $\frac{2}{9}$ & $\frac{4}{9}$\\
\hline
$ \langle \frac{3}{2},-\frac{3}{2}|$ & 0 & $\frac{2}{3}$ \\
\hline
\end{tabular}
\label{case1complete}
\end{table}

\begin{table}
\caption{Correlation space for $S_1=\frac{5}{2}$, $S_2=\frac{1}{2}$ with initial conditions and unknown variables.}

\begin{tabular}{c|c|c|c|c|c|}
%\hline
 & $\langle 2,2|$ & $\langle 2,1|$  & $\langle {2},0|$ & $\langle 2,-1|$ & $\langle 2,-2|$ \\
\hline
$\langle 3,3|$ & $\xi^{1,1}=\frac{5}{6}$ & 0 & 0 & 0 & 0\\
\hline
$\langle 3,2|$ & $\xi^{2,1}$ & $\xi^{2,2}=\frac{5}{9}$ & 0 & 0 & 0\\
\hline
$\langle 3,1|$ & $\xi^{3,1}$  & $\xi^{3,2}$ & $\xi^{3,3}$ & 0 & 0\\
\hline
$\langle 3,0|$ & 0 & $\xi^{4,2}$ &  $\xi^{4,3}$ &  $\xi^{4,2}$  & 0\\
\hline
$\langle 3,-1|$ & 0 & 0 &  $\xi^{3,3}$ & $\xi^{3,2}$ &  $\xi^{3,1}$ \\
\hline
$\langle 3,-2|$ & 0 & 0 & 0 &  $\xi^{2,2}=\frac{5}{9}$  &  $\xi^{2,1}$ \\
\hline
$\langle 3,-3|$ & 0 & 0 & 0 & 0 & $\xi^{1,1}=\frac{5}{6}$ \\
\hline
\end{tabular}
\label{case2incomplete}
\end{table}

A more complicated system of equations occurs with a $S_1=\frac{5}{2}$ $S_2=\frac{1}{2}$.  Here there is a ground state of $S_q=2$ and therefore the values of two coefficients are required to fully determine the entire correlation space matrix.

From Table \ref{coeff}, $\xi^{1,1}=\frac{5}{6}$,  $\xi^{2,2}=\frac{5}{9}$, and $\overline{\xi}=\frac{7}{18}$. Similar to the earlier case, symmetry is used to find that
\be
\xi^{1,1}=\xi^{7,5}=\frac{5}{6},~
\xi^{2,1}=\xi^{6,5},~
\xi^{3,1}=\xi^{5,5},~
\xi^{2,2}=\xi^{6,4}=\frac{5}{9},~
\xi^{3,2}=\xi^{5,4},~
\xi^{4,2}=\xi^{4,4},~
\xi^{3,3}=\xi^{5,3}
\ee

Now there are only 5 unique unknowns. The first set of equations is given from the sum of the columns. 
\be
\frac{5}{6}+\xi^{2,1}+\xi^{3,1}=\frac{7}{6}
\ee
\be
\frac{5}{9}+\xi^{3,2}+\xi^{4,2}=\frac{7}{6}
\ee
\be
2\xi^{3,3}+\xi^{4,3}=\frac{7}{6}
\ee

The next set of equations come from the sum of the rows
\be
\xi^{2,1}+\frac{5}{9}=\frac{5}{6}
\ee
\be
\xi^{3,1}+\xi^{3,2}+\xi^{3,3}=\frac{5}{6}
\ee
\be
2\xi^{4,2}+\xi^{4,3}=\frac{5}{6}
\ee

The final set of equations comes from the diagonal values, which can be written as
\be
\sum_{1}^{2S_{tot}^i+1} \xi^{i,i} = (2S_{tot}^i+1)\bar \xi_{S_1,S_2}.
\ee

This yields the equations
\be
\frac{25}{9}+\xi^{3,3}=\frac{14}{9}
\ee
\be
2\xi^{2,1}+2*\xi^{3,2}+\xi^{4,3}=\frac{14}{9}
\ee

\begin{table}
\caption{Completed correlations space for $S_1=\frac{5}{2}$, $S_2=\frac{1}{2}$}
\begin{tabular}{c|c|c|c|c|c|}
%\hline
 & $\langle 2,2|$ & $\langle 2,1|$  & $\langle {2},0|$ & $\langle 2,-1|$ & $\langle 2,-2|$ \\
\hline
$\langle 3,3|$ & $\frac{5}{6}$ &  0 & 0 & 0 & 0\\
\hline
$\langle 3,2|$ & $\frac{5}{18}$ & $\frac{5}{9}$ & 0 & 0 & 0\\
\hline
$\langle 3,1|$ & $\frac{1}{18}$  & $\frac{4}{9}$ & $\frac{1}{3}$ & 0 & 0\\
\hline
$\langle 3,0|$ & 0 & $\frac{1}{6}$ &  $\frac{1}{2}$ &  $\frac{1}{6}$  & 0\\
\hline
$\langle 3,-1|$ & 0 & 0 &  $\frac{1}{3}$ & $\frac{4}{9}$ &  $\frac{1}{18}$ \\
\hline
$\langle 3,-2|$ & 0 & 0 & 0 &  $\frac{5}{9}$  &  $\frac{5}{18}$ \\
\hline
$\langle 3,-3|$ & 0 & 0 & 0 & 0 & $\frac{5}{6}$ \\
\hline
\end{tabular}
\label{case2complete}
\end{table}